\documentclass[12pt]{article}

\begin{document}
\setcounter{footnote}{2}

\title{Large Quantum Gravity Effects:\\ Cylindrical Waves in Four
Dimensions}

\author{Mar\'{\i}a E.  Angulo \thanks{E-mail address:
M.E.Angulo-Lopez@sussex.ac.uk}
\\ {\sl Center for Theoretical Physics, University
of Sussex,}\\{\sl Brighton BN1 9QH, UK} \and Guillermo A. Mena
Marug\'{a}n\thanks{E-mail address: mena@cfmac.csic.es}\\{\sl
I.M.A.F.F., C.S.I.C., Serrano 121, 28006 Madrid, Spain}}

\date{\today}

\maketitle

\begin{abstract}

Linearly polarized cylindrical waves in four-dimensional vacuum
gravity are mathematically equivalent to rotationally symmetric
gravity coupled to a Maxwell (or Klein-Gordon) field in three
dimensions. The quantization of this latter system was performed
by Ashtekar and Pierri in a recent work. Employing that
quantization, we obtain here a complete quantum theory which
describes the four-dimensional geometry of the Einstein-Rosen
waves. In particular, we construct regularized operators to
represent the metric. It is shown that the results achieved by
Ashtekar about the existence of important quantum gravity
effects in the Einstein-Maxwell system at large distances from
the symmetry axis continue to be valid from a four-dimensional
point of view. The only significant difference is that, in order
to admit an approximate classical description in the asymptotic
region, states that are coherent in the Maxwell field need not
contain a large number of photons anymore. We also analyze the
metric fluctuations on the symmetry axis and argue that they are
generally relevant for all of the coherent states.
\end{abstract}

\renewcommand{\theequation}{\arabic{section}.\arabic{equation}}

\section{Introduction}

The quantization of spacetimes with two commuting spacelike
Killing fields has deserved intensive study during recent years
[1-11]. The main reason for this interest is the ability of this
type of spacetimes to provide a suitable framework in which one
can discuss conceptual problems and develop mathematical methods
for the quantization of general relativity. The families of
solutions with two Killing fields that have been quantized in
the literature, though simple enough as to be tractable, still
possess an infinite number of degrees of freedom, so that they
are expected to retain the field complexity that should be
present in the elusive theory of quantum gravity.

Additional motivation for the quantum analysis of spacetimes
with two commuting Killing fields comes from their possible
application to cosmology and astrophysics. Most of the
spacetimes of this kind that have been subject to quantization
can in fact be interpreted as gravitational waves propagating in
a source-free background. Particular examples are the Gowdy
cosmologies with the spatial topology of the three-torus
\cite{Me,Go}, the family of purely gravitational plane waves
\cite{MM2,MM1,Pl}, and the set of cylindrical waves in vacuum
gravity \cite{AP,KS}. Among all of them, it is the family of
cylindrical waves whose quantization has received more attention
and is probably best understood [1-7,14,15].

The pioneer of this midisuperspace approach to quantum gravity
was Kucha\v{r} \cite{Ku}. He made a preliminary discussion of
the quantum mechanics for Einstein-Rosen waves \cite{ER}, i.e.,
cylindrical gravitational waves in four dimensions with linear
polarization. The classical description of these waves is
equivalent to that corresponding to a rotationally symmetric
massless scalar field coupled to three-dimensional gravity
\cite{Ku,ABS} and, therefore, to the description of a
rotationally symmetric Einstein-Maxwell model in three
dimensions \cite{As}. Recently, a rigorous quantization of this
three-dimensional counterpart of the Einstein-Rosen model has
been carried out by Ashtekar and Pierri \cite{AP}, superseding
previous work on the subject by Allen \cite{Al}. Some technical
details concerning the self-adjointness of the metric operators
in this three-dimensional system have been revised by
Varadarajan \cite{Ma}. On the other hand, a quantum theory for
the most general family of cylindrical waves in four-dimensional
gravity, which exploits the group-theoretical properties of the
system, has been presented by Korotkin and Samtleben, although
no explicit construction has been provided for the metric
operators \cite{KS}.

An issue that has been investigated with special interest in
this context is the existence of quantum gravitational states
that can be approximated by a classical solution, or
semiclassical one if quantum matter fields are included. By
analyzing the three-dimensional theory that is obtained from the
Einstein-Rosen waves via dimensional reduction, Ashtekar has
proved that, at least in a certain sector of quantum gravity,
the semiclassical approximation may become meaningless owing to
the appearance of huge quantum gravity effects \cite{As}.
Namely, in the rotationally symmetric Einstein-Maxwell model in
three dimensions, Ashtekar has considered all coherent states of
the Maxwell field and computed their expectation values and
quantum fluctuations in the three-metric at large distances from
the center (i.e., the axis of symmetry). For coherent states
that are sharply peaked around a characteristic wave number
$k_0$, the asymptotic expectation value of the three-metric is
peaked around a classical solution if and only if
$N(e^{k_0}-1)^2\ll 1$, where $N$ is the number of photons
contained in the state. In addition, if one requires the quantum
uncertainties in the Maxwell field to be relatively small, the
semiclassical description is accurate only when $N\gg 1$
\cite{As}. Here, and in the rest of the paper, we have used
units in which $c=\hbar=8G_3=1$, $G_3$ being the gravitational
constant in three dimensions or, equivalently, the effective
Newton constant per unit length in the direction of the symmetry
axis.

The possibility of finding states in the Einstein-Maxwell system
with improved coherence in the three-metric at the expense of
increasing the dispersion in the Maxwell field was proved by
Gambini and Pullin \cite{GP}. Large quantum gravity effects
similar to those detected by Ashtekar were also found in the
rotationally symmetric Einstein-Maxwell model by employing
non-local variables \cite{DT}, and in a three-dimensional model
with toroidal symmetry \cite{Be}. Only one purely
four-dimensional gravitational system has been discussed in
which the quantum fluctuations invalidate the classical
description of the geometry: a midisuperspace model for linearly
polarized plane waves in vacuum gravity \cite{MM2}. In this
case, the huge fluctuations appear in a region where null
geodesics are focused, and not in the asymptotic region.

The aim of the present work is to revisit Ashtekar's results
about the existence of large quantum effects in cylindrical
gravity from a strictly four-dimensional point of view. The
classical equivalence of the Einstein-Rosen and the
three-dimensional Einstein-Maxwell systems does not necessarily
imply their quantum equivalence. On the other hand, since the
Einstein-Rosen model and its three-dimensional counterpart have
different metrics, all questions about the existence of quantum
states peaked around classical geometries in general relativity
should be addressed from a four-dimensional perspective. In
fact, as we will see, the Einstein-Rosen metric can be expressed
as a function of the Maxwell field and the metric in three
dimensions which is highly non-linear in the matter field. As a
consequence, coherence in the four-metric does not generally
follow from coherence in the three-metric and the field.

The plan of the paper is the following. In Sec. 2, we construct
a midisuperspace model for cylindrical waves starting from the
Hamiltonian formulation of general relativity for spacetimes
that possess two commuting spacelike Killing fields. We adopt a
gauge-fixing procedure that is similar to that introduced by
Ashtekar and Pierri in three dimensions \cite{AP}, and calculate
the reduced Hamiltonian of the model by a careful analysis of
surface terms in the gravitational action. This framework is
then particularized to the case of linearly polarized waves via
symmetry reduction. Employing the quantum theory put forward in
Ref. \cite{AP} for the rotationally symmetric three-dimensional
system, we present a complete quantum theory for the
Einstein-Rosen waves in Sec. 3. In particular, we obtain
regularized, positive operators to describe all components of
the four-metric. The behavior of these operators on the quantum
states that are coherent in the Maxwell field is discussed in
Sec. 4. We first analyze the quantum gravitational effects at
large distances of the symmetry axis, showing that the
conclusions reached by Ashtekar for the three-dimensional metric
are valid as well for the metric of the Einstein-Rosen waves,
not only qualitatively, but also quantitatively. From the point
of view of the four-metric, however, the requirement of a
classical behavior for the Maxwell field is now spurious, so
that the condition $N\gg 1$ is no longer necessary to reach an
acceptable classical approximation in the asymptotic region.
Using our four-dimensional formalism, we are also able to study
the quantum fluctuations in the metric on the symmetry axis. We
argue that these fluctuations cannot be neglected for any of the
coherent states. We summarize our results and conclude in Sec.
5. Finally, two appendices are added. In Appendix A we prove
some useful operator identities, while Appendix B contains some
calculations employed in the discussion of the metric
fluctuations.

\section{The Midisuperspace Model}
\setcounter{equation}{0}

Let us first construct a gauge-fixed midisuperspace model to
describe cylindrical waves in vacuum gravity. Since this family
of waves can be regarded as a particular class of spacetimes
that possess two commuting Killing vectors, we can start our
analysis with the Hamiltonian formulation for spacetimes of this
kind, which is discussed in Sec. 3 of Ref. \cite{MM1}. For
convenience, we adopt the notation $\{x^i\}\equiv\{Z,\theta,R\}$
($i=1,2,3$) for the spatial coordinates and assume that the two
commuting Killing vector fields are $\partial_{x^a}$ ($a=1,2$),
so that the metric is independent of $\theta$ and $Z$. In
addition, we impose that $R\geq 0$ and $\theta\in S^1$ (with
$S^1$ being the unit circle). With this terminology, $Z$ denotes
the coordinate of the symmetry axis, whereas $R$ and $\theta$
are the radial and angular coordinates on each surface of
constant $Z$ and time $t$.

The momentum constraints corresponding to the coordinates $x^a$
can be eliminated by requiring that the induced three-metric
$h_{ij}$ is block-diagonal, namely, $h_{aR}=0$ (where we have
adopted the alternative notation $h_{aR}$ instead of $h_{a3}$).
The gauge fixing is almost identical to that explained in Ref.
\cite{MM1} for the case of plane waves, and we will not repeat
details here. Apart from the different domains of definition for
the spatial coordinates, the only modification that must be
introduced concerns the system of units. In the cited paper, the
authors set $c=\hbar=4 G_3=1$, where $G_3=G/(\int dZ)$ is the
effective Newton constant per unit length. In the present work,
however, we have fixed $8G_3=1$ (to facilitate comparison of our
results with those of Ashtekar and Pierri). We can nevertheless
take account of this discrepancy by simply multiplying all
gravitational constraints in Ref. \cite{MM1} by a factor of two
and dividing the canonical momenta of the metric functions by
the same factor.

As shown in Ref. \cite{MM1}, the dynamical stability of the
gauge-fixing conditions $h_{aR}=0$ requires that
$4N\sqrt{h_{RR}}f_a=-\sqrt{{\rm det} h_{cd}} h_{ab}
(N^b)^{\prime}$, where $f_a$ are two constants (independent of
$R$) that determine the momenta of $h_{aR}$. Here, $N$ is the
lapse function, $N^i$ is the shift vector, and the prime stands
for the derivative with respect to $R$. Since the two-metric
$h_{ab}$ becomes degenerate on the symmetry axis (that we
suppose located at $R=0$), the regularity of the four-metric on
this axis implies that the constants $f_a$ must vanish. As a
consequence, we conclude that the components $N^a$ of the shift
vector are independent of the spatial coordinates, and can be
absorbed by a redefinition of $x^a$. It hence turns out that the
condition of regularity on the axis suffices to ensure that the
orbits spanned by the two Killing vectors admit orthogonal
surfaces.

The remaining momentum constraint can be eliminated in a very
similar way to that discussed at the end of Sec. 3 and the
beginning of Sec. 4 in Ref. \cite{MM1}. One only needs to change
the choice of the strictly increasing function $z_0$ that
determines the coordinate $R$. We now select $z_0=\ln{R}$. In
this way, the radial coordinate $R$ is set to coincide with the
square root of the determinant of the metric on Killing orbits.
We notice that our gauge fixing for the momentum constraint
associated with $R$ is analogous to that performed by Ashtekar
and Pierri in three dimensions \cite{AP}.

The resulting reduced system has a configuration space with
three degrees of freedom which, with the conventions of Ref.
\cite{MM1}, can be chosen as the three metric functions $v$,
$y$, and $w$. In order to adopt a notation similar to that
employed in the three-dimensional Einstein-Maxwell model
\cite{AP,As}, it is convenient to introduce the definitions
$\psi= \ln{R}-y/2$ and $\gamma=2w$. The system has still one
constraint, namely, the Hamiltonian constraint, which can now be
written \cite{MM1}
\begin{eqnarray}
{\cal H}&=&\frac{e^{(\psi-\gamma)/2}}{2R}\left[R^2(
\psi^{\prime})^2-2R\gamma^{\prime}+e^{2\psi}(v^{\prime})^2+
p_{\psi}^2+R^2e^{-2\psi}p_v^2\right]\nonumber\\ \label{hacon}
&+& e^{(\psi-\gamma)/2}p_{\gamma}(p_vv^{\prime}+p_{\psi}
\psi^{\prime}+p_{\gamma}\gamma^{\prime}-2p_{\gamma}^{\prime}).
\end{eqnarray}
The corresponding gauge freedom can be eliminated by imposing
the vanishing of the momentum canonically conjugate to $\gamma$:
$p_{\gamma}=0$. This condition is inspired by the gauge fixing
carried out in the three-dimensional counterpart of our model
\cite{AP}. It is straightforward to check that the gauge fixing
is well posed. In addition, the gauge condition is preserved by
the dynamical evolution provided that $\{p_{\gamma},\int
dR\,N{\cal H}\}\doteq -(e^{(\psi-\gamma)/2}N)^{\prime}=0$, where
the symbols $\{\;,\;\}$ and $\doteq$ denote Poisson brackets and
weak identity, respectively. Hence, the lapse function must be
of the form $N=f(t)e^{(\gamma-\psi)/2}$, with $f(t)$ being a
function of time (that can generally be absorbed by a
redefinition of $t$). We will choose this function equal to
$e^{-\gamma_{\infty}/2}$, where $\gamma_{\infty}$ is the value
of the metric function $\gamma$ when $R\rightarrow\infty$. As we
will see below, this choice guarantees that $\partial_t$ is a
unit asymptotic time translation.

On the other hand, the solution to the Hamiltonian constraint
with $p_{\gamma}=0$ is
\begin{equation}\label{gamma2}
\gamma=\frac{1}{2}\int_0^R d\bar{R}
\bar{R}\left[(\psi^{\prime})^2+\frac{p_{\psi}^2}{\bar{R}^2}+
e^{2\psi}\frac{(v^{\prime})^2}{\bar{R}^2}+e^{-2\psi}p_{v}^2\right],
\end{equation}
where we have imposed that $\gamma$ vanish at $R=0$ in order to
obtain (with suitable boundary conditions on $\psi$ and $v$) a
regular metric on the axis of symmetry. After our gauge fixing,
the line element has the expression
\begin{equation}\label{metric2}
ds^2=e^{-\psi}\left[e^{\gamma}(-e^{-\gamma_{\infty}}dt^2+dR^2)+
R^2d\theta^2\right]+e^{\psi}(dZ-vd\theta)^2.\end{equation}
Assuming as a boundary condition (see Ref. \cite{AP} for a
detailed discussion in the linearly polarized case) that the
metric functions $\psi$ and $v$ fall off sufficiently fast as
$R\rightarrow\infty$ (so that, in particular, $\gamma_{\infty}$
is finite), we get that the above metric describes an
asymptotically flat spacetime with a generally non-zero deficit
angle. In this asymptotic region, as we anticipated,
$\partial_t$ is a unit timelike vector.

The reduced model obtained in this way is free of constraints
and has only two metric degrees of freedom, described by the
variables $\psi$ and $v$. Its reduced symplectic structure is
$\Omega=\int dR ({\bf d}p_{\psi}\wedge {\bf d}\psi+{\bf d}
p_v\wedge {\bf d}v)$, where ${\bf d}$ and $\wedge$ denote,
respectively, the exterior derivative and product. The
Hamiltonian that generates the dynamics of the model, on the
other hand, can be obtained by reducing the gravitational
Hilbert-Einstein action supplemented with appropriate boundary
terms \cite{HH}. Let us explain this point in more detail. In
our gauge-fixing procedure, we have removed some of the original
degrees of freedom by expressing them in terms of the remaining
canonical variables and, possibly, of the coordinates. All the
expressions employed are in fact local, except in the very last
step of the procedure, where relation (\ref{gamma2}) has been
introduced. It is not difficult to realize then that, in our
discussion of the gauge fixing, the dynamical equations that we
have computed via Poisson brackets are actually valid in the
interior of our manifold, even though we have not explicitly
included surface terms in the Hamiltonian. This fact ensures
that our gauge fixing has been carried out consistently.
Furthermore, it then follows that the Hamiltonian of the reduced
model is actually given by the reduction of the total
Hamiltonian (including surface terms) of our original system.
Since, as we have pointed out, relation (\ref{gamma2}) is not
local, this reduced Hamiltonian may be non-trivial.

The boundary terms for the gravitational Hamiltonian have been
recently analyzed by Hawking and Hunter \cite{HH}. To apply
their results to our reduced model, let us first consider a
manifold that, on each section $\Sigma_t$ of constant time, has
a two-dimensional boundary $B_t$ which is a cylinder of radius
$R_f$ \cite{Note}. In addition, we assume that the spacetime
metric has the form (\ref{metric2}). Then, in the limit
$R_f\rightarrow\infty$ we clearly reach the family of
cylindrical waves that we want to study. Since all the
constraints have been eliminated in the process of gauge fixing
and the shift vector vanishes in Eq. (\ref{metric2}), it is not
difficult to conclude from the discussion in Ref. \cite{HH} that
the Hamiltonian of our reduced model comes exclusively from
boundary terms on $B_t$, and is given by
\begin{equation}
H=-\lim_{R_f\rightarrow \infty}2N
\sqrt{\sigma} (\kappa-\kappa_0).\end{equation}
Here, we have made $8G_3=1$, $\sigma$ is the determinant of the
two-metric induced on $B_t$, and $\kappa$ and $\kappa_0$ are the
trace of the extrinsic curvature of this metric embedded,
respectively, in $\Sigma_t$ and in a three-dimensional Minkowski
background. It is straightforward to check that
$\kappa=e^{-\gamma_{\infty}/2}/(NR_f)$ and $\kappa_0=1/R_f$,
while $\sigma=R_f^2$. Therefore, we obtain that the reduced
Hamiltonian that generates time evolution in the coordinate $t$
is $H=2(1-e^{-\gamma_{\infty}/2})$. In particular, this implies
that $\gamma_{\infty}$ is a constant of motion, because it
commutes with the Hamiltonian. So, given any classical solution,
it is possible to absorb the factor $e^{-\gamma_{\infty}}$ in
the line element by a mere rescaling of the time coordinate:
$T=e^{-\gamma_{\infty}/2}t$ (off-shell, one would have
$T=\int_0^t d\bar{t}\,e^{-\gamma_{\infty}/2}$).

Let us now particularize our considerations to the simpler case
of linearly polarized cylindrical waves. For the Einstein-Rosen
waves, we have $v=0$. We can impose this restriction as a
symmetry condition in our model. Its compatibility with the
Hamiltonian evolution leads to the secondary constraint $p_v=0$.
One can also check that there are not tertiary constraints. The
symmetry conditions $v=p_v=0$ form a pair of second-class
constraints that allow the reduction of the model by removing
two canonical degrees of freedom. The resulting system has the
following metric and symplectic structure:
\begin{eqnarray}\label{metric}
ds^2&=&e^{-\psi}[e^{\gamma}(-dT^2+dR^2)+R^2d\theta^2]+e^{\psi}dZ^2,
\\ \label{gamma} \gamma&=&\frac{1}{2}\int_0^R d\bar{R}\bar{R}
\left[(\psi^{\prime})^2+\frac{p_{\psi}^2}{\bar{R}^2}\right],\\
\label{sym}\Omega&=&\int_0^{\infty}dR\; {\bf d}p_{\psi}\wedge{\bf
d}\psi.\end{eqnarray} Note that the term between square brackets
in Eq.(\ref{metric}) is precisely the gauge-fixed metric of the
dimensionally reduced Einstein-Maxwell model discussed in Ref.
\cite{AP}. In addition, the reduced Hamiltonian coincides also
with that found by Ashtekar and Pierri in three-dimensions.
Finally, it is straightforward to see that, in terms of the time
coordinate $T$, the dynamical equations for the field $\psi$ are
exactly those satisfied by a rotationally symmetric massless
scalar field in three-dimensional Minkowski spacetime \cite{AP}.
This scalar field can be interpreted as the dual of a Maxwell
field \cite{As}. In this way, one recovers the Einstein-Maxwell
analog in three dimensions of the Einstein-Rosen waves.

\section{Quantum Theory}
\setcounter{equation}{0}

Since the field $\psi$ is a rotationally symmetric solution to
the massless Klein-Gordon equation in three dimensions that is
regular at the origin $R=0$ \cite{AP}, all classical solutions
admit the mode expansion
\begin{equation}\label{psi}
\psi(R,T)=\frac{1}{\sqrt{2}}\int_0^{\infty}dkJ_0(kR)\left[
A(k)e^{-ikT}+A^{\dagger}(k)e^{ikT}\right].\end{equation} The
constants of motion $A(k)$ and $A^{\dagger}(k)$ are complex
conjugate to each other, because $\psi$ and $J_0$ (i.e., the
zeroth-order Bessel function of the first kind) are real.
Employing the identity $2\pi J_0(kR)=\oint d\theta
e^{ikR\cos{\theta}}$, we can write the above expression in the
alternative form
\begin{equation}\label{psi3}
\psi(R,T)=\frac{1}{2\pi}\int_{I\!\!\!\,R^2}\frac{d^2k}{\sqrt{2}
\;|\vec{k}|}\left[A(|\vec{k}|)e^{i(\vec{k}\cdot\vec{x}-|\vec{k}|T)}
+A^{\dagger}(|\vec{k}|)e^{-i(\vec{k}\cdot\vec{x}-|\vec{k}|T)}\right]
,\end{equation} with $R=|\vec{x}|$. Taking then into account
that, from the Hamiltonian equations of motion,
$p_{\psi}=R\dot{\psi}$, where the overdot stands for the
derivative with respect to $T$, substitution of Eq. (\ref{psi3})
in the symplectic form leads to $\Omega=i\int_0^{\infty} {\bf
d}A^{\dagger}(k)\wedge {\bf d}A(k)$. Therefore, $A(k)$ and
$A^{\dagger}(k)$ can be understood as annihilation and creation
like variables. In addition, a trivial calculation using Eqs.
(\ref{gamma}) and (\ref{psi3}) shows that $\gamma_{\infty}$
equals the Hamiltonian of the massless scalar field \cite{AP}:
$\gamma_{\infty}=\int_0^{\infty} dk k A^{\dagger}(k)A(k)$.

Essentially, the quantization of our basic field $\psi$ can then
be carried out by introducing a Fock space in which $\psi(R,T)$
goes over to an operator-valued distribution $\hat{\psi}(R,T)$,
obtained by representing $A(k)$ and $A^{\dagger}(k)$ as standard
annihilation and creation operators \cite{AP,Ma,Al}. The Fock
space in this representation is that over the Hilbert space of
square integrable functions on the positive real axis,
$L^2(I\!\!\!\,R^+, dk)$. Using such a representation, a complete
quantization of the Einstein-Maxwell counterpart of our system
has been recently proposed \cite{AP,Ma}. Our aim in this section
is to show how the quantization put forward by Ashtekar and
Pierri in three dimensions can be employed to construct a
consistent quantum theory which fully describes the
four-dimensional metric of the Einstein-Rosen model.

As a first step towards the introduction of meaningful metric
operators, let us regularize the basic field $\hat{\psi}(R,T)$,
which is defined only as an operator-valued distribution [the
reason being that $J_0(kR)$ does not belong to
$L^2(I\!\!\!\,R^+,dk)$ for any $R\geq 0$]. Given that $J_0$ is
bounded in $I\!\!\!\,R^+$, the regularization can be achieved by
simply multiplying the factor $J_0(kR)$ in the quantum version
of Eq. (\ref{psi}) by a square integrable real function, $g\in
L^2(I\!\!\!\,R^+,dk)$,
\begin{equation}\label{psif}
\hat{\psi}(R,T|g)=\frac{1}{\sqrt{2}}\int_0^{\infty}dkJ_0(kR)g(k)\left[
\hat{A}(k)e^{-ikT}+\hat{A}^{\dagger}(k)e^{ikT}\right].
\end{equation}
This regularization can be justified from a physical point of
view, e.g., by admitting the existence of a cut-off $k_c$ in
momentum space \cite{Al}. The corresponding function $g(k)$
equals then the unity on the compact interval $[0,k_c]$ and
vanishes outside. In this sense, it is worth pointing out that
the model itself provides an energy scale, namely, $c^4/G_3$
(adopting a general system of units). Thus, a natural candidate
for $k_c$ could be $c^3/(\hbar G_3)$, which has dimensions of an
inverse length.

As an alternative motivation for the regularization, one can
just smear the operator-valued distribution
$\hat{\psi}(\vec{x},T)\equiv\hat{\psi}(R=|\vec{x}|,T)$, defined
via the quantum analog of Eq. (\ref{psi3}), with a test function
in two dimensions $f(\vec{x})$ that is also rotationally
symmetric. We assume that $f(\vec{x})$ belongs to the Schwartz
space ${\cal S}(I\!\!\!\,R^2)$ of smooth functions on the plane
with rapid decrease at infinity. In order to interpret the
smearing as an average, we further accept that $f(\vec{x})$ is
real and has a unit integral over $I\!\!\!\,R^2$. Then, a simple
calculation proves that the smeared operator $\int d^2x_0
f(\vec{x}_0)
\hat{\psi}(\vec{x}-\vec{x}_0)$ is rotationally symmetric and
can be expressed in the form (\ref{psif}), with $g(k)$ given by
\begin{equation}\label{fR}
2\pi \tilde{f}(k)=
\int_{I\!\!\!\,R^2} d^2x f(\vec{x}) e^{i\vec{k}\cdot\vec{x}}=
2\pi\int_0^{\infty} dR\; RJ_0(kR) f(R).
\end{equation}
Here, $\tilde{f}(\vec{k})$ denotes the Fourier transform of
$f(\vec{x})$ in two dimensions, and the notation $\tilde{f}(k)$
and $f(R)$ indicates that these functions depend only on
$k=|\vec{k}|$ and $R=|\vec{x}|$, respectively. The last identity
in the above equation shows that $\tilde{f}(k)$ is real; hence,
so is $g(k)$. The first identity, together with the properties
of the Fourier transform \cite{RS} and the fact that
$f(\vec{x})$ belongs to the Schwartz space, implies that
$\tilde{f}(\vec{k})\in{\cal S}(I\!\!\!\,R^2)$. Then, we have
that $g(k)$ belongs to the Hilbert space $L^2(I\!\!\!\,R^+,dk)$.
In addition, since $f(\vec{x})$ has unit integral, it follows
that $g(0)=2\pi\tilde{f}(0)=1$.

For any choice of the real function $g\in L^2(I\!\!\!\,R^+,dk)$,
the operator (\ref{psif}), with domain given by the dense
subspace of the Fock space consisting of all finite particle
vectors \cite{RS}, is symmetric and admits a self-adjoint
extension \cite{RS}, which we will denote again with the symbol
$\hat{\psi}(R,T|g)$. The standard spectral theorems ensure then
that the exponential operators $e^{\pm\hat{\psi}(R,T|g)}$ are
well-defined and positive. Besides, recalling the definition of
normal ordering and the Campbell-Baker-Hausdorff (CBH) formula
$e^{\hat{a}}
e^{\hat{b}}=e^{[\hat{a},\hat{b}]/2}e^{\hat{a}+\hat{b}}$, which
is valid for operators whose commutator is a $c$-number
\cite{MW}, we conclude
\begin{equation}\label{epsi}
:e^{\pm\hat{\psi}(R,T|g)}:=e^{-||g_R||^2/2}\;
e^{\pm\hat{\psi}(R,T|g)},\end{equation} where $||\,.\,||$
denotes the norm in $L^2(I\!\!\!\,R^+,dk)$ and
\begin{equation}\label{gR}
g_R(k)=\frac{1}{\sqrt{2}}J_0(kR)g(k).\end{equation} The diagonal
$\theta$ and $Z$ components of the four-metric can then be
represented by the positive operators
\begin{equation}\label{qmpsi}
\hat{h}_{\theta\theta}(R,T|g)=R^2 :e^{-\hat{\psi}(R,T|g)}:\;,
\;\;\;\;\;\hat{h}_{ZZ}(R,T|g)=:e^{\hat{\psi}(R,T|g)}:\end{equation}
Note that the normal ordering in these definitions guarantees
that the vacuum expectation values reproduce the classical
values of $h_{\theta\theta}$ and $h_{ZZ}$ in Minkowski
spacetime.

On the other hand, the representation of the metric function
(\ref{gamma}) by a regularized operator $:\hat{\gamma}(f_R,T):$
was discussed in Refs. \cite{AP,Ma}. The symbol $f_R$ denotes a
smearing function employed in the regularization, namely,
$f_R(r)$ is a function on the positive real axis that equals the
unity for all $r\leq R$, decreases smoothly in $[R,R+\epsilon]$
and vanishes for $r\geq R+\epsilon$, with $\epsilon> 0$ being a
certain parameter with dimensions of length \cite{Note3}. It has
been recently shown \cite{Ma} that this regularized operator has
a well-defined action on a dense subspace of the Fock space
which is contained in the set of finite particle vectors. In
that domain of definition, the operator $:\hat{\gamma}(f_R,T):$
is symmetric \cite{Ma}. As a straightforward consequence, so is
$:\hat{\gamma}(f_R,T):-\hat{\psi}(R,T|g)$ provided that the real
function $g$ belongs to $L^2(I\!\!\!\,R^+,dk)$. In addition,
Varadarajan has argued that $:\hat{\gamma}(f_R,T):$ admits a
self-adjoint extension, because it is (formally) possible to
find a conjugation \cite{RS} that leaves invariant the domain of
definition of this operator and commutes with it \cite{Ma}. In
fact, the same argument supports the existence of a self-adjoint
extension of $:\hat{\gamma}(f_R,T):-\hat{\psi}(R,T|g)$, because
it is easy to check that the considered conjugation commutes as
well with $\hat{\psi}(R,T|g)$ when $g$ is real. Using the
spectral theorem, we would then conclude that the exponential of
this self-adjoint extension,
\begin{equation}\label{RR}\hat{\Gamma}(R,T|g,f_R)\equiv e^{:\hat{
\gamma}(f_R,T):-\hat{\psi}(R,T|g)},\end{equation}
is a well-defined, positive operator.

We can then represent the remaining non-trivial components of
the four-metric (i.e., the diagonal $R$ and $T$ components) by
the operator
\begin{equation}\label{hRR}\hat{h}_{RR}(R,T|g,f_R)=
e^{-||\bar{g}_R||^2}\;\hat{\Gamma}(R,T|g_1,f_R),\end{equation}
where we have adopted the notation
\begin{equation}\label{g12}
\bar{g}_R(k)=\frac{\sqrt{e^k-1-k}}{e^k-1}g_R(k),\;\;\;\;\;
g_1(k)=\frac{k}{e^k-1}g(k),\end{equation} and used definition
(\ref{gR}). It is readily seen that the functions $\bar{g}_R$
and $g_1$ belong to $L^2(I\!\!\!\,R^+,dk)$ if so does the
function $g$. According to our discussion above, the introduced
operator should then be positive if the function $g$ is real and
square integrable on the positive axis. In our definition
(\ref{hRR}), the numerical factor $e^{-||\bar{g}_R||^2}$, as
well as the replacement of $g$ with $g_1$ as the regularization
function used in $\hat{\Gamma}$, can be understood as a
convenient choice of factor ordering. Indeed, after restoring
the dimensional constants $c$, $\hbar$, and $G_3$ in our
calculations, it is possible to check that, when
$\hbar\rightarrow 0$, the factor $e^{-||\bar{g}_R||^2}$ tends to
the unity, whereas $g_1\rightarrow g$. The selected factor
ordering is motivated by the following considerations.

In the limit $R\rightarrow\infty$, the smearing function $f_R$
tends to the unit function, and the operator
$:\hat{\gamma}(f_R,T):$ becomes
\begin{equation}\label{H0}
:\hat{\gamma}_{\infty}:=\int_{0}^{\infty}dk\;k\hat{A}^{\dagger}(k)
\hat{A}(k),\end{equation}
which is the normal ordered Hamiltonian of a rotationally
symmetric, massless scalar field in three dimensions
\cite{AP,As}. We then obtain that, in the asymptotic region
$R\rightarrow
\infty$, the purely radial component of the quantum metric is
given by
$\lim_{\bar{R}\rightarrow\infty}\hat{h}_{RR}(\bar{R},T|g,1)$. On
the other hand, it is shown in Appendix A that
\begin{equation}\label{RNO}
\hat{h}_{RR}(\bar{R},T|g,1)=e^{-\int_0^{\infty}dk g_{\bar{R}}(k)
e^{ikT}\hat{A}^{\dagger}(k)}\;e^{:\hat{\gamma}_{\infty}:}\;
e^{-\int_0^{\infty} dk
g_{\bar{R}}(k)e^{-ikT}\hat{A}(k)}.\end{equation} Therefore, our
factor ordering ensures that, at least in the asymptotic region,
the vacuum expectation value of the metric operator (\ref{hRR})
coincides with the classical value of $h_{RR}$ in Minkowski
spacetime, a value which is equal to the unity. In addition, the
factor ordering adopted is also very convenient from a practical
point of view, because, for polynomials of the operator
(\ref{RNO}), all matrix elements between coherent states of the
basic field $\psi$ are explicitly computable. For such coherent
states, one can then complete the calculation of the asymptotic
fluctuations in $\hat{h}_{RR}$. Moreover, the operator
$e^{:\hat{\gamma}_{\infty}:}$ that appears in Eq. (\ref{RNO}) is
precisely the operator employed by Ashtekar and Pierri to
represent the purely radial component of the metric in the
three-dimensional counterpart of our system \cite{AP}. As we
will see in the next section, this fact leads to a simple
relation between the coherent expectation values and
fluctuations obtained for the radial component of the asymptotic
metric in the four and three-dimensional models.

\section{Metric fluctuations}
\setcounter{equation}{0}

We are now in an adequate position to study the quantum geometry
of the model and discuss whether the conclusions obtained by
Ashtekar in three dimensions about the existence of large
quantum gravity effects in the asymptotic region generalize to
the four-dimensional model describing Einstein-Rosen waves. Like
in the analysis of Ref. \cite{As}, we will only consider quantum
states that are coherent in the basic field $\psi$. These states
show the most classical behavior that is allowed for the
fundamental field of the theory \cite{MW}. As such, they are
natural candidates in the search for states that admit an
approximate classical description of the geometry.

Given any complex function $C\in L^2(I\!\!\!\,R^+,dk)$, there
exists an associated coherent state $|C\rangle$ of unit norm,
which has the form
\begin{equation}
\label{coh}|C\rangle=e^{-||C||^{2}/2}e^{\int_0^{\infty}dk C(k)
\hat{A}^{\dagger}(k)}|0\rangle .\end{equation} Here,
$|0\rangle$ is the unique vacuum of the Fock space. For any
coherent state, the expectation value of the (regularized) field
$\hat{\psi}(R,T|g)$ coincides, at all values of $R$ and $T$,
with the classical field solution obtained by replacing the
annihilation and creation operators with the functions $C(k)$
and its complex conjugate:
\begin{equation}\label{psic}
\langle\hat{\psi}(R,T|g)\rangle_{C}=
2\int_0^{\infty}dk\,g_R(k) {\rm Re}[C(k)e^{-ikT}],
\end{equation}
with ${\rm Re}[\,.\,]$ denoting the real part. In addition,
using definitions (\ref{qmpsi}) and the CBH formula, one can
check that the coherent expectation values of the diagonal
$\theta$ and $Z$ components of the metric are also equal to the
corresponding classical expressions:
\begin{equation}\label{qmc}
\langle\hat{h}_{\theta\theta}(R,T|g)\rangle_C=
R^2 \left(\langle\hat{h}_{ZZ}(R,T|g)\rangle
_C\right)^{-1}=R^2\;e^{-\langle\hat{\psi}(R,T|g)\rangle_C}.
\end{equation}

The calculation of the expectation value of the purely radial
component of the metric is much more involved, and we will only
analyze the asymptotic case $R\rightarrow \infty$. According to
our discussion at the end of Sec. 3, this asymptotic expectation
value is equal to the limit of $\hat{h}_{RR}(\bar{R},T|g,1)$
when $\bar{R}\rightarrow \infty$. Employing Eq. (\ref{RNO}) and
the operator identities (\ref{eaeH}), one arrives at
\begin{equation}\label{hRc}
\langle\hat{h}_{RR}(\bar{R},T|g,1)\rangle_C=
\langle e^{:\hat{\gamma}_{\infty}:}\rangle_C\;
e^{-\langle\hat{\psi}(\bar{R},T|g)\rangle_C}.\end{equation}
Here, $\langle e^{:\hat{\gamma}_ {\infty}:}\rangle_C$ is
precisely the coherent expectation value obtained in three
dimensions for the diagonal $R$ component of the asymptotic
metric \cite{Note2}:
\begin{equation}\label{hR3}
\langle e^{:\hat{\gamma}_ {\infty}:}\rangle_C=e^{\int_0^{\infty}
dk\;(e^k-1)|C(k)|^2}.\end{equation} Notice that Eq. (\ref{hRc})
can be understood as the quantum counterpart of the classical
relation $h_{RR}=e^{\gamma-\psi}$ when $\gamma$ is set equal to
its asymptotic value. In fact, this non-trivial result is due to
the factor ordering adopted in Eq. (\ref{hRR}). Furthermore,
assuming that there exist strictly positive constants $k_1$ and
$\alpha$ such that the function $g(k)k^{1/2-\alpha}$ is bounded
in the interval $[0,k_1]$, we prove in Appendix B that the limit
$R\rightarrow\infty$ of the right-hand side of Eq. (\ref{psic})
vanishes. Therefore, the asymptotic expectation value of
$\hat{h}_{RR}$ in a coherent state turns out to coincide then
with $\langle e^{:\hat{\gamma}_{\infty}:}\rangle_C$. Once again,
this coincidence can be interpreted as the analog of the
classical identity $h_{RR}(R=\infty)=e^{\gamma_{\infty}}$, which
incorporates the boundary condition that $\psi$ vanish at
infinity. Taking into account that the expectation value
$\langle e^{:\hat{\gamma}_{\infty}:}\rangle_C$ equals the
classical value of $e^{\gamma_{\infty}}$ (at least) if the wave
profile $C(k)$ has negligible high-energy contributions
\cite{As}, we conclude that all coherent states in the
low-energy sector would admit an approximate classical
description of the four-dimensional geometry in the asymptotic
region provided that they have small relative fluctuations in
the metric when $R\rightarrow
\infty$.

Before continuing our analysis, let us briefly comment on the
assumption introduced above about the real function $g\in
L^2(I\!\!\!\,R^+,dk)$ employed in the regularization. The
existence of a bound in an interval starting at the origin is
clearly satisfied for the function $g$ itself (i.e., with
$\alpha=1/2$) if $g(k)$ is a cut-off in momentum space; in that
case, $g(k)\leq 1$ on the positive axis. In addition, if the
adopted regularization can be interpreted as a smooth spatial
smearing, the function $g(k)$ is bounded again on the whole
semiaxis $k\geq 0$, because $g(\vec{k})\equiv g(k=|\vec{k}|)$
given by Eq. (\ref{fR}) is a Schwartz test function in
$I\!\!\!\,R^2$. These facts strongly support our hypothesis and
show its compatibility with a wide class of feasible
regularizations.

As a first step in the calculation of the metric fluctuations in
the asymptotic region, one can check that
\begin{eqnarray}\label{flucZ}
\Xi_C\hat{h}_{\theta\theta}(\bar{R},T|g)&=&\Xi_C\hat{h}_{ZZ}(\bar{R},
T|g)=e^{||g_{\bar{R}}||^2}-1, \\ \label{fluR}
\Xi_C\hat{h}_{RR}(\bar{R},T|g,1)&=&e^{||\breve{C}||^2}e^{-\langle\hat
{\psi}(\bar{R},T|g)\rangle_{\breve{C}}}e^{||g_{\bar{R}}||^2}-1,
\end{eqnarray}
where $\breve{C}(k)=C(k)(e^k-1)$ \cite{Note1} and
$\Xi_C\hat{a}=(\langle\hat{a}^2\rangle_C/\langle\hat{a}
\rangle_C^2)-1$
is the square of the relative uncertainty in the operator
$\hat{a}$ for the coherent state $|C\rangle$. It is worth
noticing that, when $g=0$, Eq. (\ref{fluR}) reproduces the
asymptotic fluctuations in the radial component of the
three-metric studied by Ashtekar \cite{As,Note2}. In order to
deduce the value of the asymptotic fluctuations, one only needs
to take the limit $\bar{R}\rightarrow\infty$ in the above
expressions. Actually, with our assumption about the existence
of a segment to the right of the origin where the function
$g(k)k^{1/2-\alpha}$ is bounded for some choice of $\alpha>0$,
it is shown in Appendix B that the asymptotic limits of
$||g_{\bar{R}}||$ and
$\langle\hat{\psi}(\bar{R},T|g)\rangle_{\breve{C}}$ vanish.
Hence, for any of the coherent states, all metric operators
display a classical behavior in the asymptotic region, except
the operator that describes the purely radial component.
Moreover, the square of the relative uncertainty in this last
operator is given by the quantity $e^{||\breve{C}||^2}-1$, which
is precisely the value of the corresponding uncertainty in the
three-dimensional Einstein-Maxwell analog of our cylindrical
system \cite{As}.

As a straightforward consequence, it turns out that all the
results reached by Ashtekar in three dimensions about the
appearance of large quantum gravity effects apply as well to the
four-dimensional model constructed here for the Einstein-Rosen
waves. Indeed, the conclusions reached by Ashtekar are not only
qualitatively valid from a four-dimensional point of view, but
also quantitatively accurate. The only existing difference is
that, as far as the four-metric is concerned, one does not need
to demand that the relative fluctuations in the basic field
$\psi$ (and in the physical quantities associated with it, like,
e.g., the Hamiltonian) be negligible. Then, it is not necessary
that the coherent states contain a large number of elementary
excitations, a condition that is imposed in the
three-dimensional system \cite{As}. From this perspective, there
exist more coherent states that admit a classical description of
the four-metric in the asymptotic region than those that provide
a meaningful semiclassical solution to the Einstein-Maxwell
model obtained by dimensional reduction.

Summarizing, in order for the classical approximation to be
acceptable in the asymptotic region only two conditions must be
verified \cite{Note1}:
\begin{equation}\label{asy}
\int_0^{\infty} dk |C(k)|^2(e^k-1-k)\ll 1,\;\;\;\;\;
\int_0^{\infty}dk |C(k)|^2(e^k-1)^2\ll 1.
\end{equation}
The first condition ensures that the coherent expectation value
of $e^{:\hat{\gamma}_{\infty}:}$ coincides with the classical
value of $e^{\gamma_{\infty}}$. The second condition guarantees
that the asymptotic fluctuations in the radial component of the
metric are sufficiently small. In fact, the latter of these
inequalities turns out to imply the former. In particular, for a
wave profile $C(k)$ peaked around a certain wave number $k_0$
and with expected number of ``particles'' equal to $N=\int dk
|C(k)|^2$ \cite{As}, the above conditions reduce to
$N(e^{k_0}-1)^2\ll 1$.

Finally, let us notice that Eq. (\ref{flucZ}) determines the
metric fluctuations in the $\theta$ and $Z$ components at all
points of the spacetime, and not just in the asymptotic region.
It is then possible to obtain a useful estimate of those
fluctuations also on the symmetry axis $R=0$, at least for a
physically reasonable class of regularization functions $g$.
Taking into account the definition of $g_{R}$ given in Eq.
(\ref{gR}) and that the Bessel function $J_0$ equals the unity
at the origin, one can check that the square norm $||g_{R}||^2$
becomes equal to $||g||^2/2$ when one approaches the axis.
Suppose then that we further demand that the real regularization
function $g\in L^2(I\!\!\!\,R^+,dk)$ take on the constant unit
value in an interval starting at $k=0$. This interval will have
the generic form $[0,k_c]$, where $k_c$ is a positive but
otherwise arbitrary parameter. Notice that, in this case, one
can make $k_1\geq k_c$ and $\alpha=1/2$, because the function
$g$ is bounded in an interval containing $[0,k_c]$. More
importantly, according to our discussion in Sec. 3, all cut-off
functions satisfy our new condition, with the parameter $k_c$
being the cut-off introduced in momentum space. One can then
interpret every function $g$ in the considered family as a kind
of generalized cut-off. Besides, it is clear that the
regularization can still be viewed as a smooth spatial smearing
if, in addition, $g(\vec{k})\equiv g(k=|\vec{k}|)$ belongs to
${\cal S}(I\!\!\!\,R^2)$. For this class of regularization
functions, one readily obtains that $||g||^2\geq k_c$, so that,
on the axis,
\begin{equation}\label{fluax}
\Xi_C\hat{h}_{\theta\theta}=\Xi_C\hat{h}_{ZZ}\geq
e^{k_c/2}-1.\end{equation}

The relative uncertainties in the diagonal $\theta$ and $Z$
components of the metric will thus become relevant on the
symmetry axis unless $k_c\ll 1$. However, one would expect that,
in our model, a physically reasonable (generalized) cut-off
parameter $k_c$ should be at least of the order of the inverse
of the natural length scale provided by the system, i.e.,
$c^3/(\hbar G_3)\equiv k_P$ (in a general system of units). With
our conventions, $c=\hbar=8G_3=1$, and thus $k_P=8$. But, for
$k_c\geq k_P=8$, we get from Eq. (\ref{fluax}) that
$\Xi_C\hat{h}_{\theta\theta}=\Xi_C\hat{h}_{ZZ}>50$. So, quantum
gravity effects are huge on the symmetry axis for all of the
considered regularizations and, therefore, also in the limit in
which the cut-off is removed. In particular, this fact seems to
indicate that the requirement of regularity on the axis of
rotational symmetry is meaningless from a quantum mechanical
point of view.

\section{Conclusions}

We have constructed a complete quantum theory that describes the
metric of the family of Einstein-Rosen waves. This theory is
based on the quantization carried out in Ref. \cite{AP} for the
Einstein-Maxwell model obtained by the dimensional reduction of
linearly polarized cylindrical gravity.

We have started with the Hamiltonian formulation of general
relativity for spacetimes that admit two commuting spacelike
Killing vectors. Introducing suitable gauge-fixing conditions
adapted to cylindrical symmetry, we have been able to remove all
the gravitational constraints. In this way, we have arrived at a
reduced model for the most general family of cylindrical waves
in vacuum gravity. We have also calculated the symplectic
structure induced from general relativity and the Hamiltonian
that generates the time evolution. This Hamiltonian has been
computed by reducing the gravitational Einstein-Hilbert action
supplemented with appropriate surface terms. Such terms include
the contribution of the timelike boundary located at
$R\rightarrow\infty$ (where $R$ is the radial coordinate), and
have been normalized to vanish in Minkowski spacetime.

We have then imposed the requirement of linear polarization as a
symmetry condition. This has led to a reduced midisuperspace
model whose classical solutions are precisely the Einstein-Rosen
waves. The model has only one degree of freedom in configuration
space, given by a cylindrically symmetric field $\psi$, and is
indeed classically equivalent to a rotationally symmetric,
massless scalar field (dual to a Maxwell field) coupled to
three-dimensional gravity. The non-zero components of the
four-metric in our reduced model are exponentials of the basic
field $\psi$ multiplied either by trivial functions or by the
purely radial component of the three-metric in the
Einstein-Maxwell system. Employing the quantum theory proposed
in Ref. \cite{AP} for this three-dimensional model, we have then
achieved a full quantization of the metric for Einstein-Rosen
waves. Since, using a Fock space representation in which the
field $\psi$ is represented as an operator-valued distribution,
Ashtekar and Pierri had already succeeded in constructing a
(presumably \cite{Ma}) positive operator for the diagonal radial
component of the metric in three dimensions, our quantization
process has been reduced, basically, to the following two steps.
Firstly, we have regularized the field $\psi$ to reach a
well-defined operator and avoid ultraviolet divergences.
Secondly, owing to the non-linearity of the metric in $\psi$, we
have introduced a reasonable choice of factor ordering for the
metric operators.

We have also analyzed whether there exist large quantum gravity
effects in the system, as happens to be the case in the
Einstein-Maxwell counterpart of the model. We have shown that,
with the chosen factor ordering, the expectation values of the
diagonal $\theta$ and $Z$ components of the four-metric
correspond in fact to classical trajectories in all of the
coherent states of the field $\psi$. In addition, we have seen
that, like in the three-dimensional model, the asymptotic
expectation value of the radial component is that predicted by
the classical theory (semiclassical theory in three dimensions),
provided that the coherent state has negligible contributions
from the high-energy sector. In the derivation of this result,
we have introduced the very weak assumption that the function
$g\in L^2(I\!\!\!\,R ^+,dk)$, employed in the regularization of
the field $\psi$, is bounded in a certain interval around $k=0$
when multiplied by a factor of the form $k^{1/2-\alpha}$, with
$\alpha$ being a positive constant. Such an assumption is
satisfied by all cut-off regularizations, as well as by those
regularizations that can be interpreted as a smooth spatial
smearing of the field, and therefore implies no restriction in
physically relevant situations.

For such regularizations we have also computed the value of the
quantum uncertainties in the metric when one approaches the
asymptotic region. The fluctuations in the $\theta$ and $Z$
components turn out to vanish when $R\rightarrow\infty$.
Therefore, these metric operators display a classical asymptotic
behavior. As far as they are concerned, the boundary condition
that the basic field vanish asymptotically is respected quantum
mechanically in all coherent states. The asymptotic fluctuations
in the diagonal radial component, on the other hand, are exactly
the same as in the three-dimensional model \cite{As}. These
results prove the validity of the analysis carried out by
Ashtekar in three dimensions, not only qualitatively, but also
quantitatively. There is only one caveat: in order for the
four-dimensional metric to admit a classical description, it is
not needed that the physical quantities associated with the
field $\psi$ possess small relative uncertainties. The
requirement that the number of fundamental excitations contained
in the coherent state be large, necessary for a meaningful
semiclassical approximation in the three-dimensional model
\cite{As}, is no longer present. In this sense, the set of
coherent states that are peaked around a classical four-metric
in the asymptotic region is bigger than that corresponding to
acceptable semiclassical solutions in three dimensions. In
particular, the vacuum is contained only in the former of these
sets. Of course, apart from the quantum metric, one could be
interested in considering other operators related with the
four-dimensional geometry (like those which describe the
spacetime curvature). Demanding that such operators have
negligible asymptotic fluctuations might well impose further
conditions on the family of coherent states that display a
classical behavior and, perhaps, restrict again their number of
``particles''.

We have also analyzed the quantum fluctuations in the diagonal
$\theta$ and $Z$ components of the metric when one approaches
the symmetry axis. We have shown that, for all regularizations
that do not modify the mode decomposition of the field $\psi$ up
to wave numbers of the order of the natural scale
$k_P=c^3/(\hbar G_3)$, the relative uncertainties in the metric
at $R=0$ are large. In particular, they explode in the limit in
which the regularization is removed. One should then expect
significant quantum effects on the axis. It is thus unclear up
to what extent the condition of regularity of the four-geometry
on the symmetry axis is sensible from a quantum mechanical
perspective.

Although there exist other possible quantizations of our model,
the quantum theory constructed presents clear advantages. In
fact, it has been constructed in such a way that the relation
between the metric operators in three and four dimensions are as
simple and natural as possible. This fact has allowed us to
compare the physical results for the Einstein-Rosen waves with
those obtained by Ashtekar in the three-dimensional
Einstein-Maxwell model, and prove that the latter are indeed
relevant in four dimensions. Regarding the factor ordering, we
have checked that the existence of large quantum fluctuations in
the metric is rather insensitive to the operator ordering.
However, it generally affects the expectation value of the
metric in coherent states, so that, for factor orderings other
than the one selected, such value would only reproduce a
classical solution in the limit $\hbar\rightarrow 0$. Obviously,
this is one of the reasons that motivated our choice of
ordering.

On the other hand, it is worth noticing that our discussion
about the expectation value of the metric and its uncertainty in
the asymptotic region is in fact regularization independent,
apart from the more than reasonable hypothesis that the
regularization function (possibly multiplied by a factor
$k^{1/2-\alpha}$, with $\alpha>0$) be bounded in a neighborhood
of the origin of wave numbers, an assumption that, as we have
commented, involves no physical limitation in practice. Our
analysis of the metric uncertainties on the symmetry axis,
nevertheless, has been restricted to a particular (though quite
general) family of regularizations, which can be interpreted as
a generalized cut-off. In this sense, our results about the
fluctuations on the axis depend on the regularization adopted.
However, since those fluctuations are always significant when
the regularization is removed at scales below the inverse-length
parameter $k_P$, naturally provided by the system, one would
expect the existence of important quantum gravity effects on the
axis of cylindrical symmetry in all physically plausible
situations. Finally, since coherence in the basic field $\psi$
is not a requisite for the validity of the classical
approximation from a purely four-dimensional viewpoint, it would
be interesting to investigate whether the quantum fluctuations
in the four-metric can be diminished by considering other
families of quantum states, like, e.g., those analyzed by
Gambini and Pullin \cite{GP}.

\subsection*{Acknowledgments}

M. E. A. acknowledges the financial support provided by C.S.I.C.
during the completion of this work. G. A. M. M. was supported by
DGESIC under the Research Projects No. PB97-1218 and
HP1988-0040.

\section*{Appendix A}
\renewcommand{\theequation}{A.\arabic{equation}}
\setcounter{equation}{0}

In this appendix, we will prove relation (\ref{RNO}). We will
make use of the operator expansion theorem \cite{MW}
\begin{equation}\label{OET}
e^{x\hat{a}}\;\hat{b}\;e^{-x\hat{a}}=\sum_{n=0}^{\infty}
\frac{x^n}{n!}[\hat{a},\hat{b}]_{(n)}\end{equation}
and the identity \cite{MW}
\begin{equation}\label{exp}
e^{\hat{a}}\;e^{\hat{b}}\;e^{-\hat{a}}
=\exp{\left(e^{\hat{a}}\hat{b}e^{-\hat{a}}\right)}.\end{equation}
In these expressions, $\hat{a}$ and $\hat{b}$ denote two generic
operators and $[\hat{a},.]_{(n)}$ is the $n$-th application of
the commutator with $\hat{a}$. Particularizing these equations
to the case in which $\hat{b}=:\hat{\gamma}_{\infty}:$, $x=1$,
and
\begin{equation}
\hat{a}=\frac{1}{\sqrt{2}}\int_0^{\infty}\frac{dk}{k}
J_0(k\bar{R})g_1(k)\left[\hat{A}^{\dagger}(k)e^{ikT}-
\hat{A}(k)e^{-ikT}\right]
\equiv\hat{D}(\bar{R},T|g),\end{equation}
we obtain
\begin{equation}\label{step1}
\hat{h}_{RR}(\bar{R},T|g,1)=e^{-||\check{g}_{\bar{R}}||^2}\;e^{\hat{D}
(\bar{R},T|g)}\;e^{:\hat{\gamma}_{\infty}:}\;e^{-\hat{D}(\bar{R},T|
g)}.\end{equation} Here,
$\check{g}_{R}(k)=g_{R}(k)/\sqrt{e^k-1}$ and we have employed
definitions (\ref{gR}) and (\ref{g12}). On the other hand, a
repeated application of the operator expansion theorem to
calculate the commutator of $e^{:\hat{\gamma}_{\infty}:}$,
firstly with the smeared version of the creation and
annihilation operators, and then with their exponentials, leads
to
\begin{eqnarray}e^{:\hat{\gamma}_{\infty}:}\;
e^{\int_0^{\infty}
dkf(k)\hat{A}^{\dagger}(k)}&=&e^{\int_0^{\infty}
dkf(k)e^k\hat{A}^{\dagger}(k)}\;e^{:\hat
{\gamma}_{\infty}:},\nonumber\\
\label{eaeH} e^{\int_0^{\infty}
dkf(k)\hat{A}(k)}\;e^{:\hat{\gamma}_{\infty}:}&=&e^{:\hat
{\gamma}_{\infty}:}\;e^{\int_0^{\infty}
dkf(k)e^k\hat{A}(k)}.\end{eqnarray} Using these relations,
together with the CBH formula, one can readily check that the
right-hand sides of Eqs. (\ref{RNO}) and (\ref{step1}) coincide.

\section*{Appendix B}
\renewcommand{\theequation}{B.\arabic{equation}}
\setcounter{equation}{0}

We want to prove that the expectation value
$\langle\hat{\psi}(R,T|g)\rangle_C$ and the norm $||g_R||$
vanish in the asymptotic limit $R\rightarrow\infty$ if the
functions $C$ and $g$ belong to the Hilbert space
$L^2(I\!\!\!\,R^+,dk)$ and, for some choice of positive constant
$\alpha$, the function $g(k)k^{1/2-\alpha}$ is bounded in an
interval of the form $[0,k_1]$. Here, $k_1$ is a strictly
positive number and $g_{R}(k)=J_0(kR)g(k)/\sqrt{2}$. In fact, we
only need to show that $||g_{R}||$ vanishes in the asymptotic
region, because, using Eq. (\ref{psic}), the triangle inequality
for complex numbers, and the Schwarz inequality on
$L^2(I\!\!\!\,R^+,dk)$, one gets
\begin{equation}
\left|\langle\hat{\psi}(R,T|g)\rangle_C\right|\leq 2 ||C||
\;||g_R||.\end{equation}
Obviously, the same arguments apply to the value of
$\langle\hat{\psi}(R,T|g)\rangle_{\breve{C}}$ appearing in Eq.
(\ref{fluR}). Let us then write
\begin{equation}
||g_R||^2=\frac{1}{2}\int_0^{k_1} dk
J^2_0(kR)|g(k)|^2\,+\frac{1}{2}\int_{k_1}^{\infty}dk
J^2_0(kR)|g(k)|^2.\end{equation} The second term on the
right-hand side vanishes when $R\rightarrow\infty$ because, in
that limit, $J^2_0(kR)$ tends to zero uniformly in $k\in
[k_1,\infty)$, with $k_1>0$. As for the first term, let $G$ be
the upper bound of $|g(k)k^{1/2-\alpha}|$ in $[0,k_1]$. Then
\begin{equation}\frac{1}{2}
\int_0^{k_1}dkJ^2_0(kR)|g(k)|^2\leq\frac{G^2}{2R^{2\alpha}}
\int_0^{k_1R}\frac{dk}{k^{1-2\alpha}}
J_0^2(k).\end{equation} Recalling that $\alpha>0$ and
$J_0(k)\approx
\cos{(k-\pi/4)}\sqrt{2/(\pi k)}$ (up to subdominant terms) for
$k\gg 1$, one can finally show that the limit of the above
expression when $R\rightarrow\infty$ is zero.


\begin{thebibliography}{27}

\bibitem{AP} A.  Ashtekar and M.  Pierri, {\it J. Math. Phys.}
{\bf 37}, 6250 (1996).

\bibitem{KS} D.  Korotkin and H.  Samtleben, {\it Phys. Rev.
Lett.} {\bf 80}, 14 (1998).

\bibitem{Ma} M. Varadarajan, {\it Class. Quantum Grav.} {\bf
17}, 189 (2000).

\bibitem{As} A. Ashtekar, {\it Phys. Rev. Lett.} {\bf 77}, 4864
(1996).

\bibitem{GP} R. Gambini and J. Pullin, {\it Mod. Phys. Lett.}
{\bf A12}, 2407 (1997).

\bibitem{DT} A. E. Dom\'{\i}nguez and M. H. Tiglio, {\it Phys. Rev.}
{\bf D60}, 064001 (1999).

\bibitem{CMN} J. Cruz, A. Mikovi\'c, and J. Navarro-Salas,
{\it Phys. Lett.} {\bf B437}, 273 (1998).

\bibitem{Me} G. A. Mena Marug\'{a}n, {\it Phys. Rev.} {\bf D56},
908 (1997).

\bibitem{MM2} G. A. Mena Marug\'{a}n and M. Montejo, {\it Phys. Rev.}
{\bf D61} (in press), gr-qc/9906101.

\bibitem{MM1} G. A. Mena Marug\'{a}n and M. Montejo, {\it Phys.
Rev.} {\bf D58}, 104017 (1998).

\bibitem{Be} C. Beetle, {\it Adv. Theor. Math. Phys.} {\bf 2},
471 (1998).

\bibitem{Go} B. K. Berger, {\it Ann. Phys.} (N.Y.) {\bf 83},
458 (1974); {\bf 156}, 155 (1984); V. Husain and L. Smolin, {\it
Nucl. Phys.} {\bf B327}, 205 (1989).

\bibitem{Pl} D. E. Neville, {\it Class. Quantum Grav.} {\bf 10},
2223 (1993); {\it Phys. Rev.} {\bf D55}, 766 (1997); {\bf D55},
2069 (1977); {\bf D56}, 3485 (1997); {\bf D57}, 986 (1998); R.
Borissov, {\it ibid.} {\bf D49}, 923 (1994).

\bibitem{Ku} K. Kucha\v{r}, {\it Phys. Rev.} {\bf D4}, 955 (1971).

\bibitem{Al} M. Allen, {\it Class. Quantum Grav.} {\bf 4}, 149
(1987).

\bibitem{ER} A. Einstein and N. Rosen, {\it J. Franklin Inst.}
{\bf 223}, 43 (1937).

\bibitem{ABS} A. Ashtekar, J. Bi\v{c}\'{a}k, and B. G. Schmidt,
{\it Phys. Rev.} {\bf D55}, 669 (1997); {\bf D55}, 687 (1997).

\bibitem{HH} S. W. Hawking and C. J. Hunter, {\it Class.
Quantum Grav.} {\bf 13}, 2735 (1996).

\bibitem{Note} We may also assume the existence of
boundaries $B_t$ given by an initial and a final surface of
constant $Z$. These boundaries, however, have a vanishing
contribution to the Hamiltonian, beacuse the metric is
independent of $Z$ and the normals to the considered surfaces
have opposite orientation.

\bibitem{RS} M. Reed and B. Simon, {\it Methods of Modern
Mathematical Physics II: Fourier Analysis, Self Adjointness}
(Academic Press, New York, 1975).

\bibitem{MW} L. Mandel and E. Wolf, {\it Optical Coherence and
Quantum Optics} (Cambridge University Press, Cambridge, England,
1995).

\bibitem{Note3} We partly modify the form of $f_R$ given in
\cite{AP,Ma} to avoid problems with the definition of this
function when $R\rightarrow 0$.

\bibitem{Note2} With our notation, the constant $G$ appearing in
the purely radial component of the metric in Ref. \cite{As}
corresponds in fact to $8G_3$.

\bibitem{Note1} To avoid divergences, we restrict our
considerations to the dense subspace of coherent states for
which $\breve{C}(k)=C(k)(e^k-1)$ belongs to
$L^2(I\!\!\!\,R^+,dk)$. This restriction was implicitly made in
Ref. \cite{As}, because the fluctuations in the three-metric
involve the exponential of $||\breve{C}||^2$.



\end{thebibliography}
\end{document}